\documentclass[twocolumn,numberedappendix]{emulateapj}

\usepackage{amsmath}
\usepackage{natbib}

\newcommand{\pasa}{Publications of the ASA}
\submitted{Submitted to the Astronomical Journal, September 2, 2014}

\newcommand{\Decision}{$(c_1 \ldots c_5) = ( 0, \, 10, \, 1,\, 0.3,\, 2.5)$}
\newcommand{\NumberHumanComponents}{118}
\newcommand{\NumberAGDComponents}{120}
\newcommand{\InitialAlphaOneOnePhase}{$ \log \alpha_1 = 3.00$}  
\newcommand{\alphaOneOnePhase}       {$ \log \alpha_1 = 1.29$} 
\newcommand{\AccOnePhase}            {$0.78$}                        
\newcommand{\AccOnePhaseOOS}         {$0.71$}                       
\newcommand{\InitialAlphaOneTwoPhase}{$ \log \alpha_1 = 1.3$}   
\newcommand{\InitialAlphaTwoTwoPhase}{$ \log \alpha_2 = 3.0$}   
\newcommand{\alphaOneTwoPhase}       {$ \log \alpha_1 = 1.12$} 
\newcommand{\alphaTwoTwoPhase}       {$ \log \alpha_2 = 2.73$} 
\newcommand{\AccTwoPhase}            {$0.81$}                       
\newcommand{\AccTwoPhaseOOS}         {$0.79$}                       
\newcommand{\NBetterTwoPhase}        {$\Delta N = +0.1$}  
\newcommand{\rmsPcntBetterTwoPhase}  {$f_{\rm rms} = -2.2\% $}           
\newcommand{\NBetterOnePhase}        {$\Delta N  = -0.14$} 
\newcommand{\rmsPcntBetterOnePhase}  {$f_{\rm rms} = +29\% $}

\begin{document}

\title{Autonomous Gaussian Decomposition}

\author{Robert R. Lindner\altaffilmark{1$\dagger$},  
        Carlos Vera-Ciro\altaffilmark{1},             
        Claire E. Murray\altaffilmark{1},             
        Sne\v{z}ana Stanimirovi\'{c}\altaffilmark{1}, 
        Brian L. Babler\altaffilmark{1}, \\              
        Carl Heiles\altaffilmark{2},                  
        Patrick Hennebelle\altaffilmark{3},           
        W. M. Goss\altaffilmark{4}, and              
        John Dickey\altaffilmark{5}}                 
                 
\altaffiltext{1}{Department of Astronomy, 
                 University of Wisconsin, 
                 Madison, WI 53706, USA}
\altaffiltext{2}{Radio Astronomy Lab, UC Berkeley, 
                 601 Campbell Hall, Berkeley, 
                 CA 94720, USA}
\altaffiltext{3}{Laboratoire AIM, Paris-Saclay, 
                 CEA/IRFU/SAp-CNRS-Universit\'{e} Paris Diderot, 
                 91191 Gif-sur Yvette Cedex, France}
\altaffiltext{4}{National Radio Astronomy Observatory, 
                 P.O. Box O, 1003 Lopezville, Socorro, 
                 NM, 87801, USA}
\altaffiltext{5}{University of Tasmania, School of Maths and Physics, 
                 Private Bag 37, Hobart, TAS 7001, Australia}

\altaffiltext{$\dagger$}{rlindner@astro.wisc.edu}

\begin{abstract}
We present a new algorithm, named Autonomous Gaussian 
Decomposition (AGD), for automatically decomposing spectra 
into Gaussian components.  AGD uses derivative 
spectroscopy and machine learning to provide optimized 
guesses for the number of Gaussian components in the data,
and also their 
locations, widths, and amplitudes.    
We test AGD and find that it produces results comparable to 
human-derived solutions on 21\,cm absorption 
spectra from the 21\,cm SPectral line Observations of Neutral 
Gas with the EVLA (21-SPONGE) survey.
We use AGD with Monte Carlo methods to derive the H{\sc i} line completeness
as a function of peak optical depth and velocity width for
the 21-SPONGE data, and also show that the results of AGD are
stable against varying observational noise intensity.
The autonomy and computational efficiency of the method over traditional manual Gaussian fits 
allow for truly unbiased comparisons between observations 
and simulations, and for the ability to scale up and 
interpret the very large data volumes 
from the upcoming Square Kilometer Array and 
pathfinder telescopes.
\end{abstract}

\section{Introduction: 21\,cm Gaussian Fits}

Neutral hydrogen (H{\sc i}) is the raw fuel for star formation 
in galaxies, and an important ingredient in 
understanding galaxy formation and evolution through cosmic
time.
In the interstellar medium (ISM) of the Milky Way, H{\sc i} 
is predicted to exist in
two thermally stable states: the cold neutral medium (CNM) with 
temperature between 40--$200\,\rm K$, and the warm neutral medium (WNM) with
temperature between 4100--$8800\,\rm K$ \citep{field1969,mckee1977,wolfire2003}.

The 21\,cm hyperfine transition of H{\sc i} is a convenient tracer
of H{\sc i}.  One technique
for measuring the excitation temperature of H{\sc i} 
is to fit 21\,cm emission and absorption data 
to a collection of independent 
iso-thermal Gaussian components. With
this technique,  H{\sc i} spin temperatures have been measured 
in the range of 
$\sim 10$--$3000\,\rm K$ \citep[e.g., ][]{mebold1982,
kalberla1985,dickey1978,crovisier1980,heiles2001,
dickey2003,heiles2003,begum2010,roy2013}.
Although H{\sc i} surveys have detected hundreds of 
components with temperatures consistent with the 
predictions for the CNM, few have been detected with temperatures
consistent with those of the WNM.
The absence of WNM-temperature gas is surprising because 
the WNM contains $\sim 50\%$ of the mass in the 
neutral ISM \citep{draine2010}.  Surveys also find 
a significant fraction of thermally {\em unstable} gas 
(with temperature between $\sim 200$--$4100\,\rm K$), 
up to 47\% of detections in \citet{heiles2003}.  Although the missing WNM 
could be explained in terms of sub-thermal excitation 
of the 21\,cm line in low density environments 
\citep[e.g., ][]{liszt2001}, recent results from 
\citet{murray2014} point instead toward a lack of 
absorption observations with enough sensitivity to detect 
WNM-temperature gas, which has an absorption strength 
$\sim 100\times$ less than CNM-temperature gas.
Additionally, numerical simulations have shown
that magnetic fields and 
non-equilibrium physics like bulk flows and turbulence 
can affect the expected relative fractions of WNM, CNM, and 
intermediate temperature (unstable) gas
\citep{audit2005,heitsch2005,maclow2005,clark2012, hennebelle2014}, although
observational data cannot yet distinguish between 
these scenarios.
The main reason it has been difficult to make progress 
in understanding the neutral ISM is that
observational surveys have sample sizes of only 
10--100 sightlines, leaving large statistical errors
in the measurements of the H{\sc i} spin temperature distribution.

The Square Kilometer Array\footnote{www.skatelescope.org} 
(SKA) and its pathfinder telescopes,
the Australian SKA 
Pathfinder\footnote{www.atnf.csiro.au/projects/askap.index.html}
(ASKAP), the recently expanded Karl G. Jansky Very
Large Array\footnote{https://science.nrao.edu/facilities/vla},
and MeerKAT\footnote{www.ska.ac.za/meerkat},
will push radio astrophysics into a new
era of ``big spectral data" by providing scientists with 
millions of high resolution, high-sensitivity radio
emission and absorption spectra probing lines of 
sight through the Milky Way and neighboring
galaxies.  This infusion of data 
promises to revolutionize our understanding of 
the neutral ISM.
However, these new data will bring new 
challenges in data interpretation. 
Modelling a 21\,cm emission or absorption spectrum 
as a superposition of $N$ independent Gaussian 
components requires solving a non-linear 
optimization problem with $3\,N$ parameters.
Because Gaussian functions do not form an 
orthogonal basis (solutions are not unique), 
the parameter space is non-convex (contains local 
optima instead of a single, global optimum), 
and therefore the final solutions sensitively depend 
on the initial guesses of the components' positions, 
widths, and amplitudes, and especially on the total 
number of components.  To minimize the chances of 
getting stuck in local optima during model fitting, researchers choose 
the initial parameter guesses to lie as close 
to the global optimum as possible.  In previous and current surveys, 
these initial guesses are provided manually,
effectively using the pattern-recognition skills of 
humans to identify the individual components within the 
blended spectra.  This manual selection process is time 
consuming and subjective, rendering it 
ineffective for the large data volumes in the 
SKA era.  Automatic line finding and Gaussian 
decomposition algorithms can solve these problems.

However, the available algorithms for automatic line detection
are either unlikely to scale to the data volumes of
SKA, or lack the flexibility to fit complex spectra.
The Bayesian line finder by \citet{allison2011} searches
parameter space using the nested sampling algorithm 
\citep{skilling2004}, and uses Bayesian inference to 
discover the optimal number of spectral components.  
However, it has only been applied to simple spectra 
with few components, and has not been tested on complex 
Galactic 21\,cm data.  Procedural algorithms like  
those of \citet{haud2000} or \citet{nidever2008}
iteratively add, subtract, or merge components
based on the effects these decisions have on the resulting 
residuals of least-squares fits, and have been used to 
interpret large datasets from, e.g., the Leiden-Argentina-Bonn (LAB)
All-Sky H{\sc i} survey \citep{kalberla2005}.  
However, the initial parameters for each fit 
are adopted from previous solutions in adjacent 
sky positions, thereby limiting the use these 
algorithms to densely-sampled emission surveys.  
Topology-based algorithms like Clumpfind \citep{williams1994} 
and Duchamp \citep{whiting2012} are too limited for efficient 
Gaussian decomposition because they can only detect components 
that are strong enough to produce local maxima in their 
spectra, do not allow components to overlap, and do not 
provide estimates of spectral shape.
Similarly, GaussClumps \citep{stutzki1990} only locates
strong components that produce local optima in 3D space.
While the above algorithms operate successfully in
the data for which they were designed, they are 
not suited for rapid objective decomposition of
millions of complex absorption spectra.

In this paper, we present a new algorithm, 
called Autonomous Gaussian Decomposition (AGD), 
which uses computer vision and machine learning to 
quickly provide optimized guesses for the initial 
parameters of a multi-component Gaussian model.  
AGD allows for the interpretation of large volumes 
of spectral data and for the ability to objectively compare 
observations to numerical simulations in a statistically robust way.  
While the development of AGD was motivated by radio astrophysics, 
specifically the 21-SPONGE 
survey \citep[][Murray et al. 2014b, in prep.]{murray2014}, 
the algorithm can be used to search for one-dimensional 
Gaussian (or any other single-peaked spectral profile)-shaped
components in any data set.

In Section \ref{s:agd}, we explain the algorithm; in Section
\ref{s:code} we describe the Python implementation of AGD
called GaussPy; in Section \ref{s:performance}, we discuss 
AGD's performance in decomposing real 21\,cm absorption 
spectra; and in Section \ref{s-conclude}, we present a 
discussion of results and conclusions.\\

\section{Autonomous Gaussian Decomposition}
\label{s:agd}

\begin{figure}
    \centering
    \includegraphics[scale=0.45]{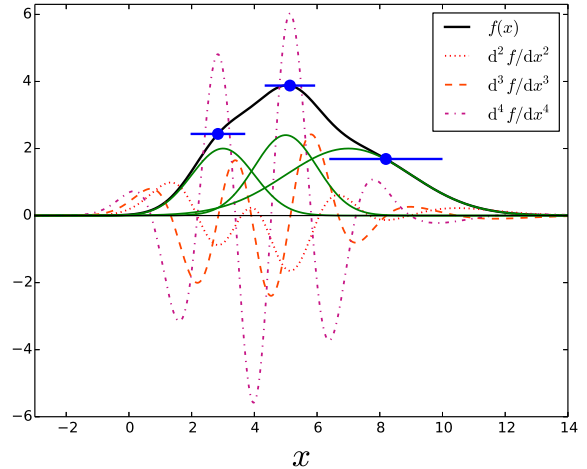}    
    \caption{Derivative spectroscopy example. 
    The green and black solid lines show the individual components
    and total signal, respectively, for a noise-free  
    spectrum consisting of three Gaussian components.
    Over-plotted are the $2^{\rm nd}$ (red dot), 
    third $3^{\rm rd}$ (orange dash), and 
    $4^{\rm th}$ (pink dot-dash) numerical derivatives.
    The locations in the data satisfying the conditions from 
    Equations \ref{e-c1}-\ref{e-c4} are identified with blue circles, 
    with blue line segments showing the guessed  $\pm 1\sigma$ 
    widths from Equation \ref{e-widths}.  The positions and widths
    indicated by the blue circle and line segments represent the
    guesses that AGD would produce for this example spectrum.}
    \label{f-example-dspec}
\end{figure}
AGD approaches the
problem of Gaussian decomposition by focusing on the
task of choosing the parameters' initial guesses,
where human input has been most needed in the 
past.  By quickly producing 
high quality initial 
guesses, most of the work in interpreting the spectrum 
has been done, and the resulting least-squares 
fit converges quickly to the global optimum. 

In the following, $x$ and $f(x)$ represent an example
spectrum. For example, $x$ might have units of frequency 
and $f(x)$ units of flux density.  Where relevant,
all one-dimensional variables are to be interpreted 
as column vectors.  The variables 
$a$, $\sigma$, and $\mu$ represent the amplitude, 
``$1\sigma$'' width (hereafter referred to as 
just the ``width''), and position of a Gaussian function $G$
according to
\begin{equation}
    G(x; a,\mu,\sigma) = a\,e^{-(x-\mu)^2/2\sigma^2}.
\end{equation}

\subsection{Derivative spectroscopy}
Derivative spectroscopy is the technique of analyzing
a spectrum's derivatives to gain understanding
about the data.  It is used in computer vision applications
because derivatives can respond to shapes in images 
like gradients, curvature, and edges.  It has a 
long history of use in biochemistry 
\citep[see, e.g., ][]{fell1983}, and has been recently 
used to analyze the spectral features of H{\sc i} 
self-absorption in two Galactic molecular clouds \citep{krco2008}.

AGD uses derivative spectroscopy to decide how 
many Gaussian components a spectrum contains, and also
to decide where they are located.  
The algorithm places one guess at the location of 
every {\em local minimum of negative curvature} in the
data, where the curvature of 
$f(x)$ is defined as the second derivative, 
${\rm d}^2 f(x)/{\rm d}x^2$.  
This criterion finds ``bumps'' in the data, and
has the sensitivity to detect weak and blended 
components.
Mathematically, this condition corresponds to 
locations in the data which satisfy the four conditions: 
\begin{align}
    & f > \epsilon_0 \label{e-c1} \\
    & {\rm d}^2f/{\rm d}x^2 < 0 \label{e-c2} \\
    & {\rm d}^3f/{\rm d}x^3 = 0 \label{e-c3}\\
    & {\rm d}^4f/{\rm d}x^4 > 0\label{e-c4}.
\end{align}
In ideal, noise-free data, we could set $\epsilon_0=0$; however, 
observational noise produces random curvature fluctuations and
a signal threshold should be applied to avoid placing guesses in
signal-free regions.
Equation \ref{e-c2} enforces that the curvature is negative, 
while Equations \ref{e-c3}--\ref{e-c4} ensure the location is
a local minimum of the curvature.
The $N$ discrete values of $x$ satisfying 
Equations \ref{e-c1}--\ref{e-c4} serve as the guesses for the 
component positions $\mu_n$ where $n\in \{1 \ldots N\}$.
Figure \ref{f-example-dspec} shows an example of applying 
Equations \ref{e-c1}--\ref{e-c4} to find the component
locations in an ideal noise-free spectrum.

Next, AGD guesses the components' widths
by exploiting the relation between a component's width
and the maximum of its second derivative. 
For an isolated component, the peak of the 
$2^{\rm nd}$ derivative is located at $x=\mu$,
and has a value of
\begin{align}
    \frac{{\rm d}^2}{{\rm d}x^2} G(x;a,\mu,\sigma) \bigg|_{x=\mu} & =\notag \\
    \frac{a}{\sigma^4} e^{-(x-\mu)^2/2\sigma^2}\left[(x-\mu)^2-\sigma^2\right]\bigg|_{x=\mu} & = -\frac{a}{\sigma^2}.
\end{align}
AGD applies this single-component solution
to provide estimates for the widths of all $n$ 
components $\sigma_n$ by approximating
$a\approx f(x)$ to obtain
\begin{equation}
    \label{e-widths}
    \sigma_n^2 = f(x) \left(  \frac{{\rm d}^2 f(x)}{{\rm d}x^2} \right)^{-1}\Bigg|_{x=\mu_n}.
\end{equation}

Finally, AGD guesses the components' amplitudes, $a_n$.  
Naive estimates for the amplitudes of the $N$ 
components are simply the values of 
the original data evaluated at the component positions.  However, if the components are highly
blended, then the naive guesses can significantly over estimate
the true amplitudes.  AGD compensates for this overestimate by
attempting to ``de-blend'' the amplitude guesses using
the information in the already-produced position and 
width guesses (See Appendix \ref{a-deblend} for 
details on the deblending process).
\begin{figure}
    \centering
    \includegraphics[scale=0.8]{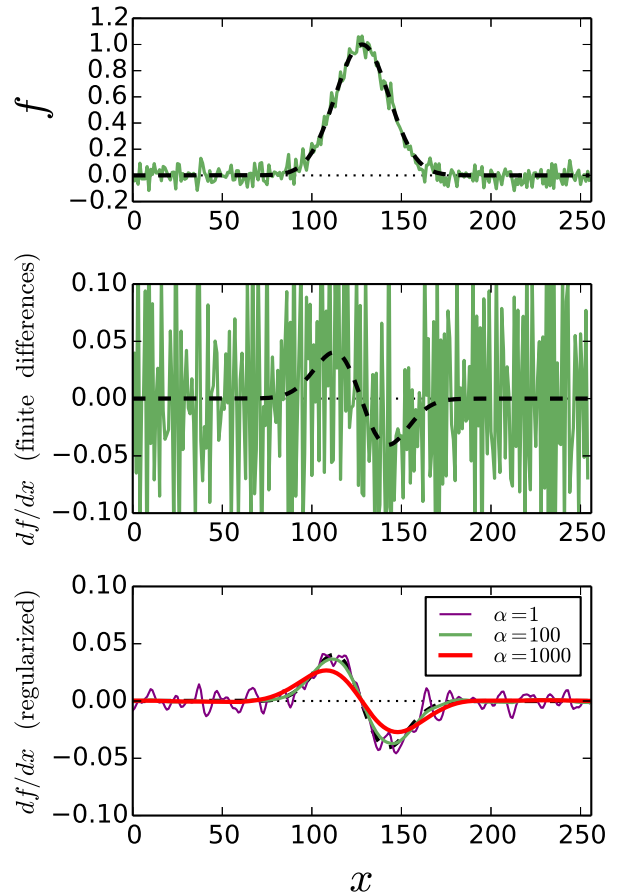}
    \caption{Regularized numerical derivatives.
    {\em Top panel}: Example spectrum is shown in green for a 
    single Gaussian component (dashed line) 
    with $a=1$, $\sigma=15$, and $\mu=128$ in Gaussian-distributed 
    noise with $\rm RMS =0.05$. {\em Middle panel:} The black dashed line shows
    the ideal derivative of the underlying Gaussian function, and
    the green line shows the results of a finite-difference-based
    numerical derivative applied to the green data from the top panel. 
    The amplified noise makes it impossible to
    locate local optima reliably.  {\em Bottom panel:}  Regularized
    derivatives (Section \ref{s-reg}) of the green data from the top panel
    using different values of the regularization parameter $\alpha$.
    Larger or smaller values of $\alpha$ trade smoothness for 
    data fidelity, respectively.\\}
    \label{f-example}
\end{figure}

\subsection{Regularized Differentiation}
\label{s-reg}

In order to identify components in $f(x)$ using
Equations \ref{e-c1}--\ref{e-c4}, the derivatives 
of $f(x)$ must be accurate and smoothly varying. Any noise in 
the derivatives of the spectra will produce spurious 
component guesses.
Computing derivatives using finite-difference
techniques greatly amplifies noise in
the data, thereby rendering finite-difference techniques 
unusable for our needs of computing derivatives 
up to the fourth order.  
We regularize\footnote{Regularization techniques are 
also used in e.g., the maximum entropy method of 
synthesis image deconvolution \citep{taylor1999}, 
and in gravitational lens image inversion 
\citep{wallington1994}.} the differentiation process
using Tikhonov regularization \citep{tikhonov1963},
where the derivative is fit to the data under the
constraint that it remains smooth by following the
technique presented in \citet{vogel2002} and
\citet{chartrand2011}.

\begin{figure}
    \centering
    \includegraphics[scale=0.5]{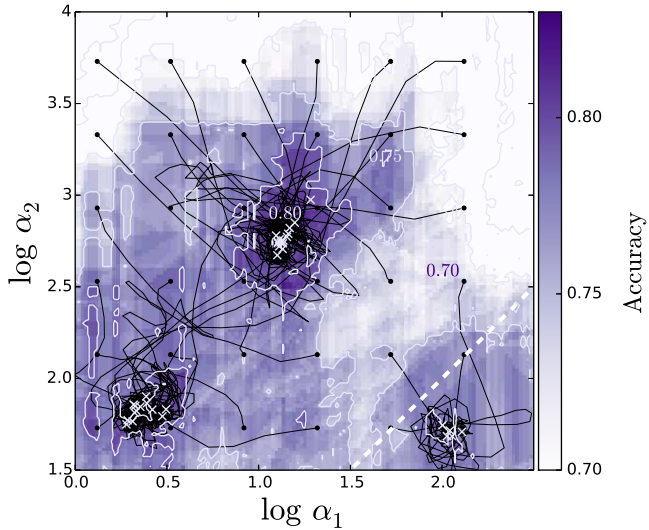}
    \caption{Training AGD using gradient descent.  
    Starting locations, tracks, and convergence locations 
    of the parameters ($\log \alpha_1, \log \alpha_2$) during 
    AGD's two-phase training process (Appendix \ref{a-gd})
    are represented by black circles, black lines, and white ``X''s, 
    respectively. The dashed white line marks the 
    $\log \alpha_1 = \log \alpha_2$ boundary.  
    Tracks that begin too far away from 
    the global best solution (\alphaOneTwoPhase{}, \alphaTwoTwoPhase{)} can converge into local optima 
    with lower resulting accuracy.  Multiple training runs with different
    starting positions are therefore recommended to find the global optimum.  
    Additionally,   physical considerations like the expected width 
    of components can help guide the choice of starting value.
    The background image shows a densely sampled representation of the
    underlying parameter space, and was generated using the 
    HTCondor cluster at the University of Wisconsin's Center for
    High-Throughput computing.}
    \label{f-tracks}
\end{figure}
We define the regularized 
derivative of the data as $u_{\rm min}=\underset{u}{\rm arg\,min}(R[u])$, where
\begin{equation}
    \label{e-TV}
    \begin{aligned}
        & R[u] = \alpha \int \sqrt{ \left(D_x u\right)^2+\beta^2 } + 
        \int |A_x u-f|^2.
    \end{aligned}
\end{equation}
The derivative operator $D_x u = {\rm d}u/{\rm d}x$ and the 
anti-derivative operator $A_x u = \int_0^x u \,{\rm d}x$.
The first term on the right-hand-side (RHS) of 
Equation \ref{e-TV} is the 
regularization term, and is
an approximation to Total-Variation (TV) regularization.
When $\beta$ is zero, this term becomes the
$L_1$ norm of ${\rm d}u/{\rm d}x$, pushing $u$ to be piecewise 
constant.  When $\beta$ is $>0$, the regularization term 
behaves like the $L_2$ norm of ${\rm d}u/{\rm d}x$, constraining $u$ 
instead to be 
smoothly varying.  
To produce smoothly-varying
solutions for our derivatives, we (a) set $\beta = 0.1$, and
(b)  rescale the bin widths to
unity and peak-normalize the data; these scale factors are
remembered and reapplied when optimization has completed.

The second term of the RHS of Equation \ref{e-TV} is the
data fidelity term, enforcing that the integral 
of $u$ closely follows the data $f$.  
The parameter $\alpha$ controls the relative balance 
between smoothness and data fidelity in the solution, i.e., between 
variance and bias. When $\alpha=0$, $u_{\rm min}$ is equal 
to the finite difference derivative.  

Figure \ref{f-example} displays how the regularization 
parameter $\alpha$ 
affects the shape of the resulting regularized derivative of a 
single Gaussian component within Gaussian distributed noise.  
Larger values of $\alpha$ effectively ignore variations in the
data on increasingly larger spatial scales.  Because of the
large range that $\alpha$ can span, we hereafter refer to the
regularization parameter as $\log_{10} \alpha \equiv \log \alpha$.

\subsection{Choosing $\log \alpha$ with machine learning}
\label{s-ml}
In supervised machine learning, the computer is given a collection
input/output pairs, known as a training set, and then ``learns'' a general
rule for mapping inputs to outputs.  After this ``training'' 
process is completed, the 
algorithm can be used to predict the output values for new inputs 
\citep[see, e.g., ][]{bishop2006,zeljko2014}.

The regularization procedure of Section \ref{s-reg} allows us 
to take smooth derivatives at the expense of introducing the
free parameter $\log \alpha$, which controls the degree of regularization.
Supervised machine learning is used to train AGD 
and pick the optimal value of $\log \alpha$ which maximizes the 
accuracy of component guesses on a training set of 
spectra with known decompositions.  One can obtain the 
training set by manually decomposing a subset of the data,
or by generating new synthetic spectra using components 
that are drawn from the same distribution as the science 
data.  In the latter case, there is a risk that the training data 
are different from the science data, but also the benefit that
the decompositions are guaranteed to be ``correct'' while the
manual decompositions are not.

Given $N_g$ component guesses
$\{a_i^g, \mu_i^g, \sigma_i^g\}_{i=1}^{N_g} \equiv g_\alpha$, 
produced by running AGD with fixed $\log\alpha$ 
on data containing $N_t$ true 
components $\{a_i^t, \mu_i^t, \sigma_i^t\}_{i=1}^{N_t} \equiv t$,
the ``accuracy'' $\mathcal{A}$ of the guesses is defined using
the balanced F-score.  The balanced F-score is 
a measure of classification accuracy that depends
on both precision (fraction of guesses that are correct) 
and recall (fraction of true components that were found), thus
penalizing component guesses which are incorrect, missing, or 
spurious.
The accuracy is given by
\begin{equation}
    \label{e-acc}
    \mathcal{A}( g_\alpha, t) = \frac{2 N_{c}}{N_{g} + N_{t}}
    \equiv  \mathcal{A}( {\log \alpha}),
\end{equation}
where $N_c$ represents the number of  ``correct'' guesses.
We consider a single guessed component 
($a^g$, $\sigma^g$, $\mu^g$) to be a ``correct''
match to a true component ($a^t$, $\sigma^t$, $\mu^t$) if 
its amplitude, position, and width are all within
the limits specified by the following equations:
\begin{align}
c_1 < \frac{a^g}{a^t} < c_2 \label{e-d1}\\
\left| \frac{\mu^g-\mu^t}{\sigma^t}\right| < c_3 \label{e-d2} \\
c_4 < \frac{\sigma^g}{\sigma^t} < c_5 \label{e-d3}.
\end{align}
The analysis in Section \ref{s:performance} 
uses \Decision{}.
The final solution is least sensitive to the initial
amplitudes, so we choose the values $c_1$ and $c_2$ to 
bracket a large relative range; it is more sensitive to
the guessed widths, so we chose a narrower relative range
in $c_4$ and $c_5$; finally, we find that the positions 
are the most important parameters for fitting the data
in the end, motivating the relatively strict value
of $c_3$.
We impose the additional restriction that matches between guessed 
and true components must be one-to-one, and therefore match
consideration proceeds in order of decreasing amplitude.

The optimal value of $\log \alpha$ 
is that which maximizes the accuracy (Equation \ref{e-acc}) 
between AGD's guessed components and the true answers
in the training data.  This non-linear optimization
process is performed using gradient descent and is described
in detail in Appendix \ref{a-gd}.

\section{GaussPy: The Python implementation of AGD}
\label{s:code}
GaussPy is the name of our
Python\footnote{GaussPy uses the NumPy \citep{walt2011}, 
SciPy \citep{jones2001} and matplotlib 
\citep{hunter2007} packages}/C implementation of the AGD algorithm.
This lightweight python module is easy to
deploy on high-throughput computing solutions like 
HTCondor\footnote{http://research.cs.wisc.edu/htcondor/} (see 
Figure \ref{f-tracks}) or 
Hadoop\footnote{http://hadoop.apache.org/}/MapReduce \citep{dean2004}, allowing 
for rapid decomposition of very large spectral datasets, 
e.g., the spectral data products of the SKA.
AGD may also be suitable for deployment on the 
Scalable Source Finding Framework \citep{westerlund2014}.
GaussPy is maintained by the author and will be publicly available
through the Python Package Index\footnote{https://pypi.python.org/pypi}
upon publication of this manuscript.
\begin{figure*}
    \centering
    \begin{tabular}{cc}
        \includegraphics[scale=0.9]{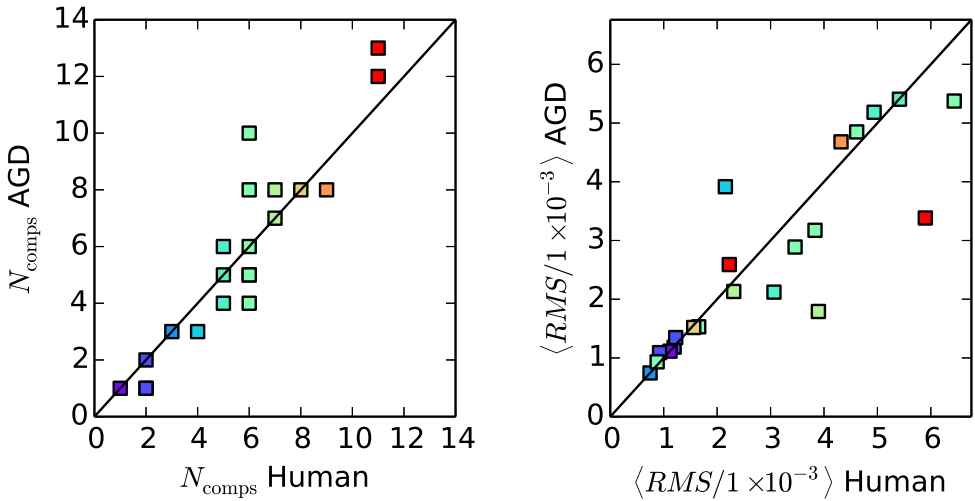}
    \end{tabular}
    \caption{AGD vs. human results for the number of Gaussian components 
    ({\em left}) and RMS residuals ({\em right}) for guess + final fit Gaussian 
    decompositions to 21 spectra from the 21-SPONGE survey.  The color
    scale represents the number of human-selected components 
    (corresponding to the $x$-axis of the left panel).}
    \label{f-correlations}
\end{figure*}

The AGD algorithm as explained in Section \ref{s:agd}
is optimized for finding components spanning only a 
modest range in width. 
This is the cost we pay for the ability to compute smooth 
derivatives using regularization.  In order to search for 
Gaussian components on widely different scales, e.g., to 
search for components with widths near 
$1$--$3\,\rm km\,s^{-1}$ and $20$-$30\,\rm km\,s^{-1}$ in
the same spectra, we can iteratively apply 
AGD to search for components with widths at each of these scales.  
This capability is included in GaussPy and is referred
to as ``two-phase'' decomposition (for details, see 
Appendix \ref{a-two-phase}).

GaussPy uses AGD to produce the initial guesses for parameters
in a multi-component Gaussian fit, and also carries out
the final least-squares fit on the data.
In this final optimization, GaussPy uses unconstrained 
minimization with the Levenberg-Marquardt 
\citep{levenberg1944} algorithm, which has been used in 
previous surveys \citep[e.g., ][]{heiles2003}.  If negative 
amplitudes are found, then the fit is remade using  the
Limited-memory bound constrained
Broyden-Fletcher-Goldfarb-Shanno (L-BFGS-B) \citep{zhu1997,nocedal2006}
algorithm for which we can enforce the constraint that all amplitudes 
be non-negative.

In GaussPy, we minimize the
functional $R[u]$ (Equation \ref{e-TV}) 
using the quasi-Newton algorithm ``BFGS2'' from
the GNU Scientific 
Library\footnote{http://www.gnu.org/software/gsl/} \texttt{Multimin} package
and achieve computation-time scalings of
${\mathcal O} \left(n^{1.95}\right)$, where $n$ is the
number of channels in the data, and ${\mathcal O} \left(\alpha^{-0.4}\right)$.
The relative 
scaling between $\log \alpha$ and the minimum preserved scale in the data 
is found numerically to be approximately
\begin{equation}
\label{e-scale}
    \delta_{\rm chan} \simeq 3.7 \times \, 1.8^{\log \alpha},
\end{equation}
where $\delta_{\rm chan}$ is the spatial scale in channels (see Figure \ref{f-example}).
By plugging in an estimate of the expected component widths to
Equation \ref{e-scale}, one obtains a rough estimate of the
appropriate regularization parameter $\log \alpha$.  
However, to find the value 
which maximizes the accuracy of the decompositions, one should solve 
for $\log \alpha$ using the machine learning technique of 
Section \ref{s-ml}.

\section{Performance: 21\,cm absorption}
\label{s:performance}
We test AGD by comparing its results
to human-derived answers for 21 spectra from
the 21\,cm SPectral line Observations of Neutral 
Gas with the EVLA (21-SPONGE) survey 
(Murray et al. 2014b, in prep.).  21-SPONGE spectra
cover a velocity range from  $-100$ 
to $ + 100 \,\rm km\,s^{-1}$, 
tracing Galactic H{\sc i} gas.  21-SPONGE's $21\,\rm cm$
absorption spectra are among the most sensitive ever 
observed with typical optical-depth root-mean-square 
(RMS) sensitivities of 
$\sigma_\tau\lesssim 10^{-3}$ per $0.4\,\rm km\,s^{-1}$ 
channel (Murray et al. 2014b, in prep.).    This 
combination of sensitivity and spectral resolution will 
stay among the best obtainable through the SKA era.
The survey data come natively in units
of fractional absorption ($I/I_{0}$), we transform the
data into optical depth units ($\tau = -{\rm ln}(I/I_{0})$)
for the AGD analysis because only in $\tau$-space 
will a single component produce a single peak 
in curvature (i.e, strong absorption signals will 
produce {\em dual} peaks in the curvature of $I/I_0$).

We begin by constructing the training data set, which
is based on independent 21\,cm absorption 
observations from the  
Millennium Arecibo 21 Centimeter Absorption-Line Survey 
\citep{heiles2003}.  We produce 20 synthetic spectra by 
 randomly-selecting Gaussian components from 
the \citet{heiles2003} catalog. The 
number of components per spectrum is chosen to be a 
uniform random integer between the mean value (three) 
and the maximum value (eight) from the observations.  
Only components with peak optical depth 
$\tau < 3.0$ are included in the training data because 
beyond this, the absorption signal saturates 
and the component properties are poorly constrained.
We next add Gaussian-distributed noise with 
$\rm RMS = 10^{-2}$ per $0.4\,\rm km\,s^{-1}$ channel to 
the spectra (in observed $I/I_0$ space) to mimic real observational noise from the
Millenium survey \citep{heiles2003}, and re-sample
the data at $0.1\,\rm km\,s^{-1}/channel$ to avoid aliasing the
narrowest components (with FWHMs of $\sim 1\,\rm km\,s^{-1}$) in 
the training set.  We set the global threshold, 
(parameter $\epsilon_0$ in Equation \ref{e-c1}), to be
$5\times$ the RMS for individual spectra.
\begin{figure}
    \centering
    \includegraphics[scale=0.65]{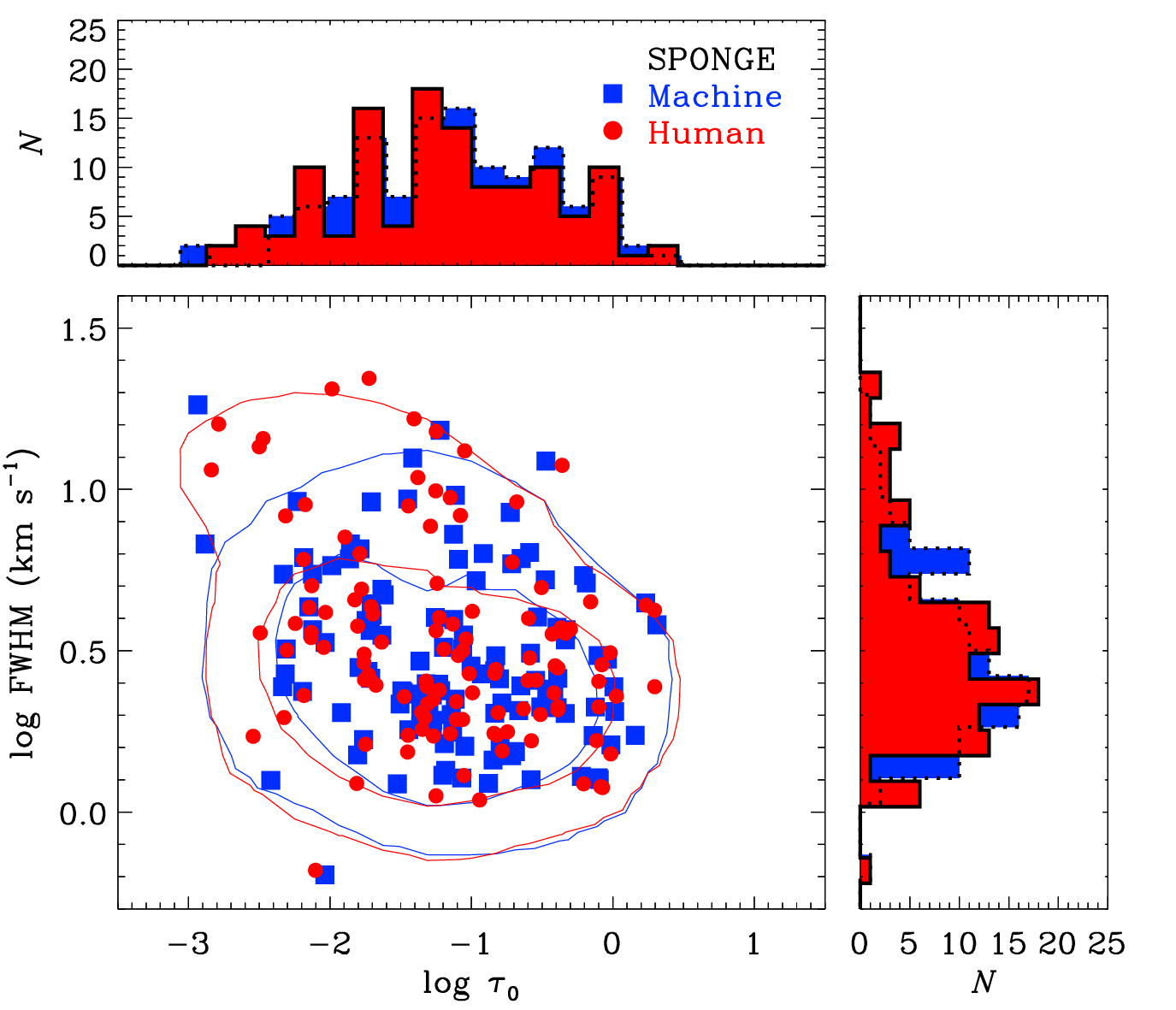}
    \caption{AGD (squares) vs. Human (circles) Gaussian decomposition results 
    for 21-SPONGE spectra.  The central panel shows peak optical 
    depth ($\tau_0$) and velocity FWHM for each recovered Gaussian 
    component.  The contours represent $68\%$ and $95\%$ containment 
    regions.  The side panels show marginalized histograms of peak 
    optical depth (top) and velocity FWHM (right) for AGD 
    (dashed) and Human (solid) results.  There are 
    \NumberHumanComponents{} Human-detected  components, and 
    \NumberAGDComponents{} AGD-detected components in the 21 spectra.}
    \label{f-21sponge}
\end{figure}

We next train AGD for both one- and two-phase decompositions
and compare their performances.
For one-phase AGD we use the initial 
value \InitialAlphaOneOnePhase{} and AGD converged 
to \alphaOneOnePhase{}.  The resulting accuracy 
was \AccOnePhase{} on the training data, 
and \AccOnePhaseOOS{} on an independent test-set of 100 
newly-generated (out-of-sample) synthetic spectra.
Testing the performance on similar but
independent out-of-sample ``test'' data prevents 
against ``over-fitting'' the training data.
For two-phase AGD, we use initial values of \InitialAlphaOneTwoPhase{}
and \InitialAlphaTwoTwoPhase{} 
and AGD converged to  \alphaOneTwoPhase{}, \alphaTwoTwoPhase{}, returning
\AccTwoPhase{} on the training data and \AccTwoPhaseOOS{}
on the independent test data from above.
Figure \ref{f-tracks} shows the convergence tracks of
$(\log \alpha_1, \log \alpha_2)$ when the two-phase training process
is initialized with different initial values for $\log \alpha_1$ 
and $\log \alpha_2$.
The $\log \alpha$ values between one- and two-phase decompositions 
generally follow the trend 
$\log \alpha_1^{\rm two\,phase} < \log \alpha^{\rm one\,phase} < \log \alpha_2^{\rm two\,phase}$, 
and this property can be used to help choose initial values during training.

We next apply the trained algorithm to the 21-SPONGE data.
We find that two-phase AGD performs better than one-phase in
decomposing the 21-SPONGE data, which 
contain absorption signatures from
two distinct populations of ISM clouds: cold clouds with narrow
absorption features and warm clouds with broad absorption features.
We compare the performance of AGD to human decompositions using
the average difference in the number of modelled 
components $\Delta N$:
\begin{equation}
    \Delta N = \left< N_{\rm AGD} - N_{\rm Human} \right>,
\end{equation}
and
the average fractional change in the residual RMS,  $f_{\rm rms}$:
\begin{equation}
    f_{\rm rms} = \left< \frac{\rm RMS_{AGD} - RMS_{Human}}{\rm RMS_{Human}} \right>.
\end{equation}
We find that \NBetterOnePhase{} and \rmsPcntBetterOnePhase{} 
for one-phase AGD and \NBetterTwoPhase{} and 
\rmsPcntBetterTwoPhase{} for two phase AGD.  
Both one-phase
and two-phase AGD guessed comparable numbers of components, 
but two-phase AGD resulted in lower residual 
errors compared to human-decomposed spectra, consistent with 
two-phase AGD's higher accuracy (i.e., \AccTwoPhaseOOS{} vs. \AccOnePhaseOOS{}, 
for two and one-phase AGD, respectively). 
A comparison between the resulting number of components and
RMS residuals between two-phase AGD and human results for
the individual spectra is shown in Figure \ref{f-correlations}.

\begin{figure*} 
    \centering
    \includegraphics[scale=1.0, trim=1 1 1 1,clip]{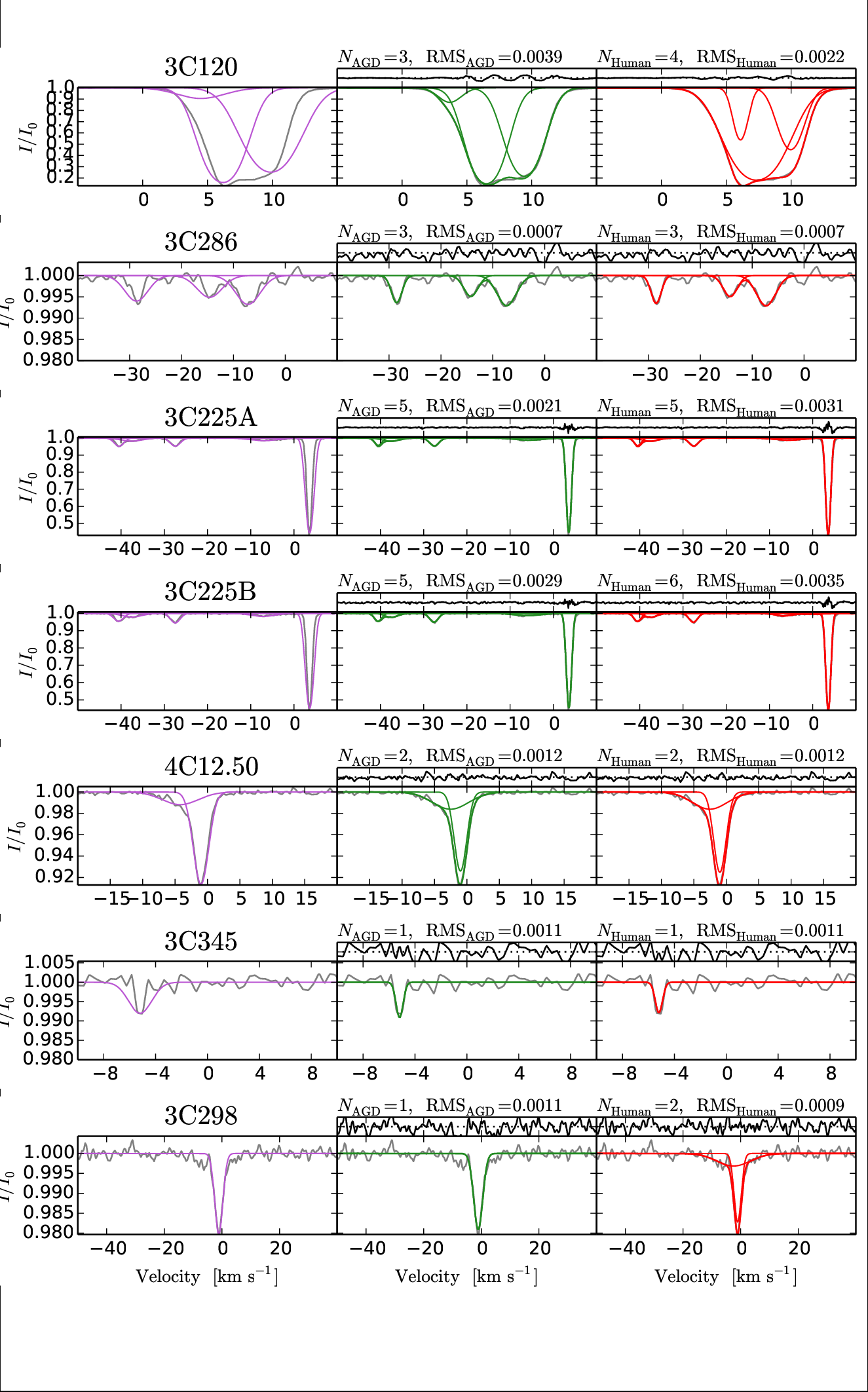}
    \caption{AGD vs. human Gaussian decompositions for 21-SPONGE absorption spectra.  
    The left panels show AGD's initial guesses (purple), the center
    panels show the resulting best-fit Gaussian components (thin green)
    and total model (thick green) found by initializing a least-squares fit
    with these initial guesses, and the right panel shows the
    human-derived best-fit components (thin red; Murray et al. 2014b, in prep) 
    and resulting models  (thick red).  The
    residual errors between the best-fit total models and the data are shown above
    the center (AGD) and right panels (Human).  The
    number of components in each fit, the source names, and the 
    residual RMS values are indicated in the panels.}
\label{fig:spectra}
\end{figure*}
\addtocounter{figure}{-1}
\begin{figure*} 
    \centering
    \includegraphics[scale=1.0, trim=1 1 1 1,clip]{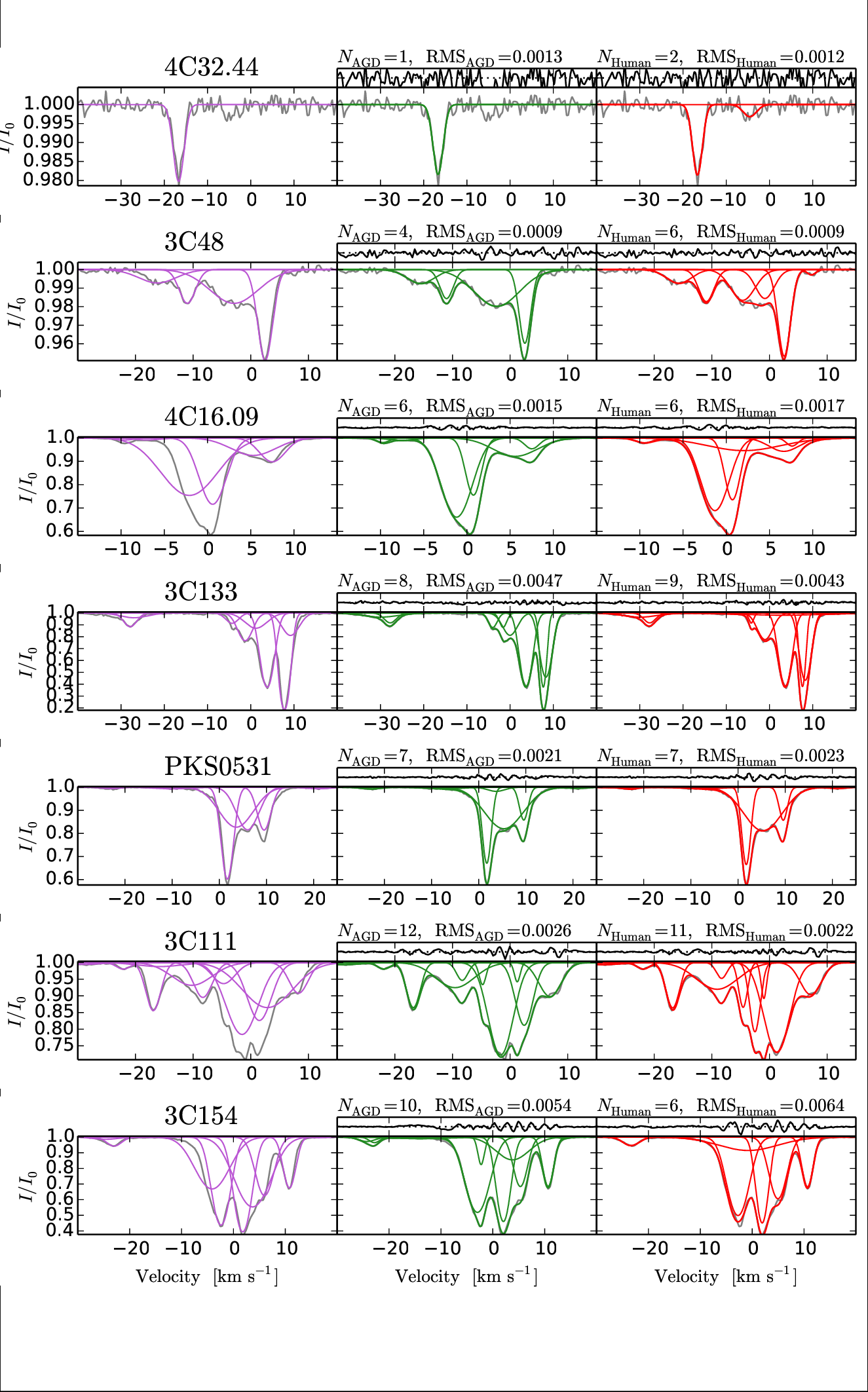}
    \caption{(continued)}
\end{figure*}
\addtocounter{figure}{-1}
\begin{figure*} 
    \centering
    \includegraphics[scale=1.0, trim=1 1 1 1,clip]{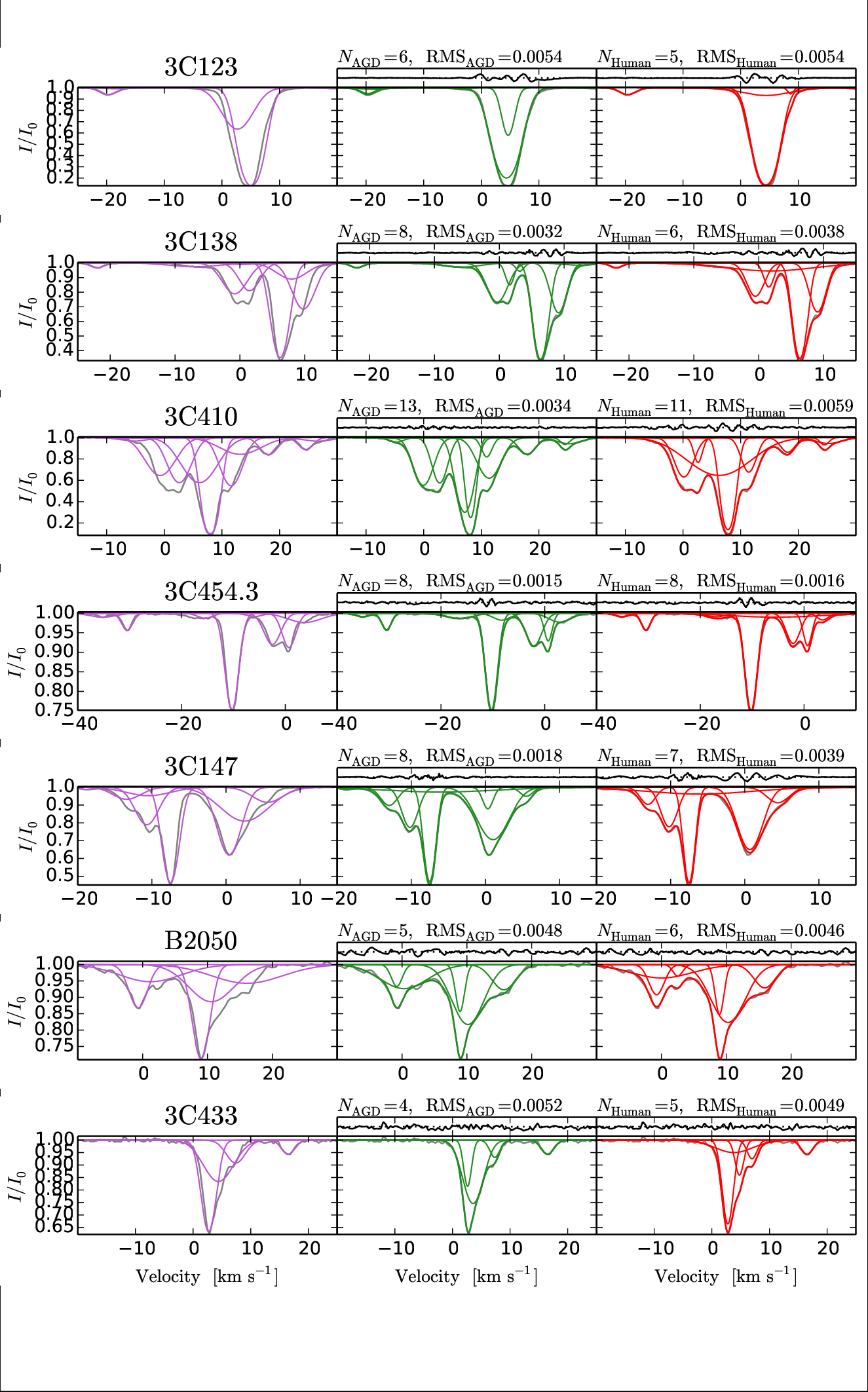}
    \caption{(continued)}
\end{figure*}
In Figure \ref{f-21sponge}, we show a scatter plot
of the best-fit FWHMs and peak amplitudes for all AGD and 
human-derived Gaussian components for the 21-SPONGE data.
There are 118 and 120 components detected by AGD and 
human, respectively.  We performed a 2-sample Kolmogorov-Smirnov 
test on the amplitudes, FWHMs, and derived equivalent widths (${\rm EW}=\int \tau(v){\rm d}v$) of the 
resulting components from AGD vs. human results and find 
that in each case, the AGD and human distributions are
consistent with being drawn from identical 
distributions.  Thus, AGD results are statistically 
indistinguishable to human-derived decompositions in terms 
of the numbers on components, the residual RMS values, and 
the component shapes.
Figure \ref{fig:spectra} shows the AGD guesses, AGD-best fits, 
and human-derived best fits for all 21 spectra in
our data set.

\subsection{Component completeness}

Observational noise can scatter the measured signals
of weak spectral lines below a survey's 
detection threshold, effectively modifying 
the measured component distribution by a ``completeness'' function.
The effect of completeness needs to be taken into
account in order to make high-precision comparisons between
the measured distributions of H{\sc i} absorption/emission profiles 
and the predictions of physical models.
AGD's speed and autonomy allows for easy reconstruction of
a survey's completeness function, and this information 
can be used to correct the number counts of observed 
line components so that one can infer the 
true component distribution to lower column densities.

\begin{figure}
    \centering
        \includegraphics[scale=0.45]{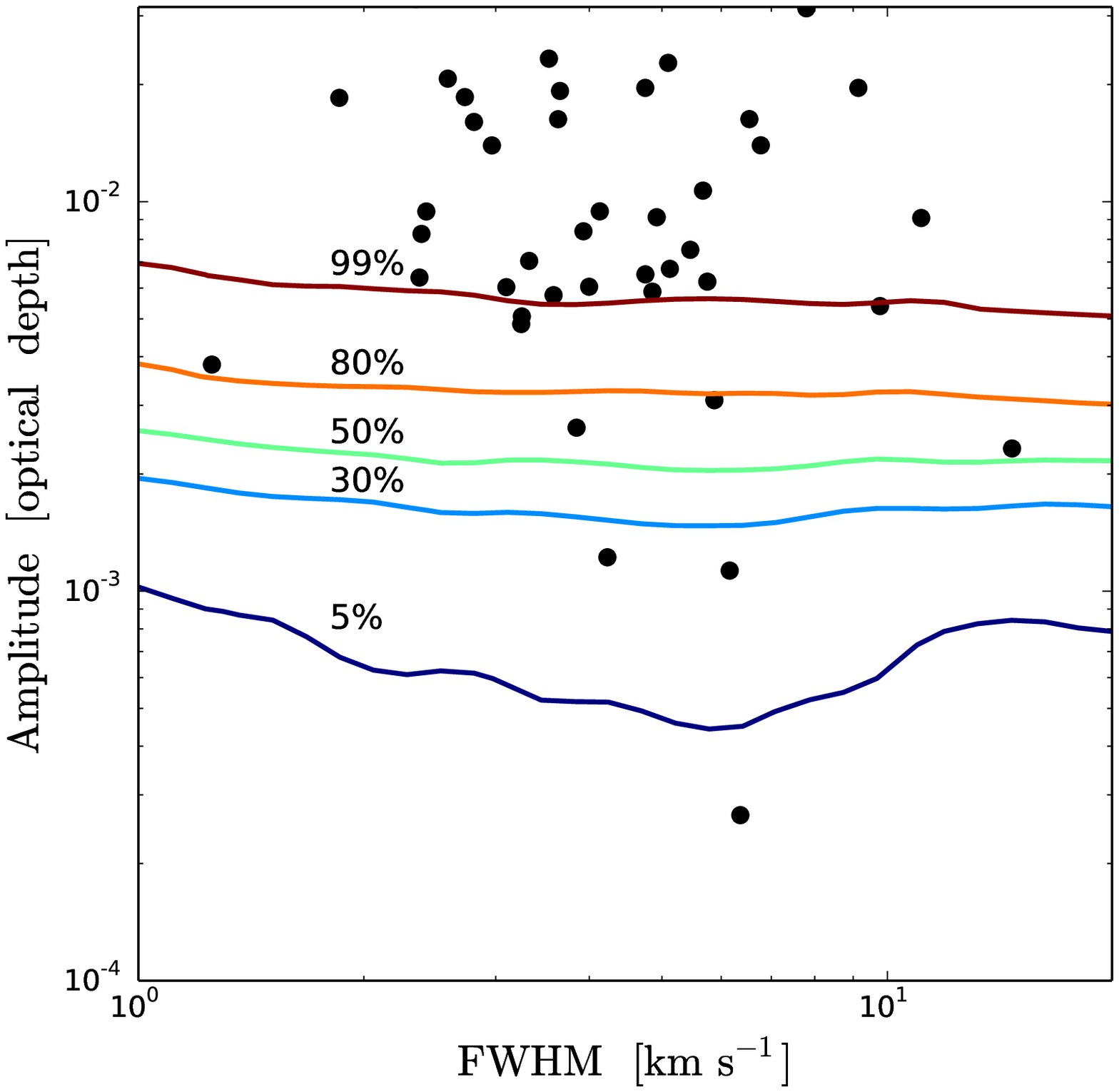}
    \caption{Completeness as a function of component amplitude
    and FWHM.  The labelled contours represent the 
    probability of detecting a component of
    a given shape, and is equal to the ratio of successful detections
    in a Monte Carlo simulation where single components were injected
    one at a time into spectra containing realistic noise of
    $\rm RMS = 10^{-3}$ per $0.4\,\rm km\,s^{-1}$ channel.  The black
    circles represent the 21-SPONGE detections from AGD. Detected components
    with amplitudes $\tau_0 > 7\times 10^{-3}$ have $\simeq 100\%$ completeness.\\}
    \label{f-completeness}
\end{figure}
We demonstrate this procedure by measuring AGD's line 
completeness of 21-SPONGE H{\sc i} absorption profiles 
as a function of amplitude and velocity width using 
a Monte Carlo simulation.  We inject a single Gaussian 
component with fixed parameters into sythetic spectra
containing realistic observational noise 
($\rm RMS = 10^{-3}$ per $0.4\,\rm km\,s^{-1}$ channel) 
and then run AGD to measure the completeness, which we define as the 
fraction of successfully-detected components out of 50
trials.  AGD's completeness function for the 21-SPONGE data 
is shown in Figure \ref{f-completeness}.
AGD obtains $\simeq 100\%$ completeness for 
components with $\rm FWHM > 1\,\rm km\,s^{-1}$ and 
$\tau_0 > 7\times 10^{-3}$.

\subsection{Robustness to varying observational noise}

\begin{figure}
    \centering
    \includegraphics[scale=0.53]{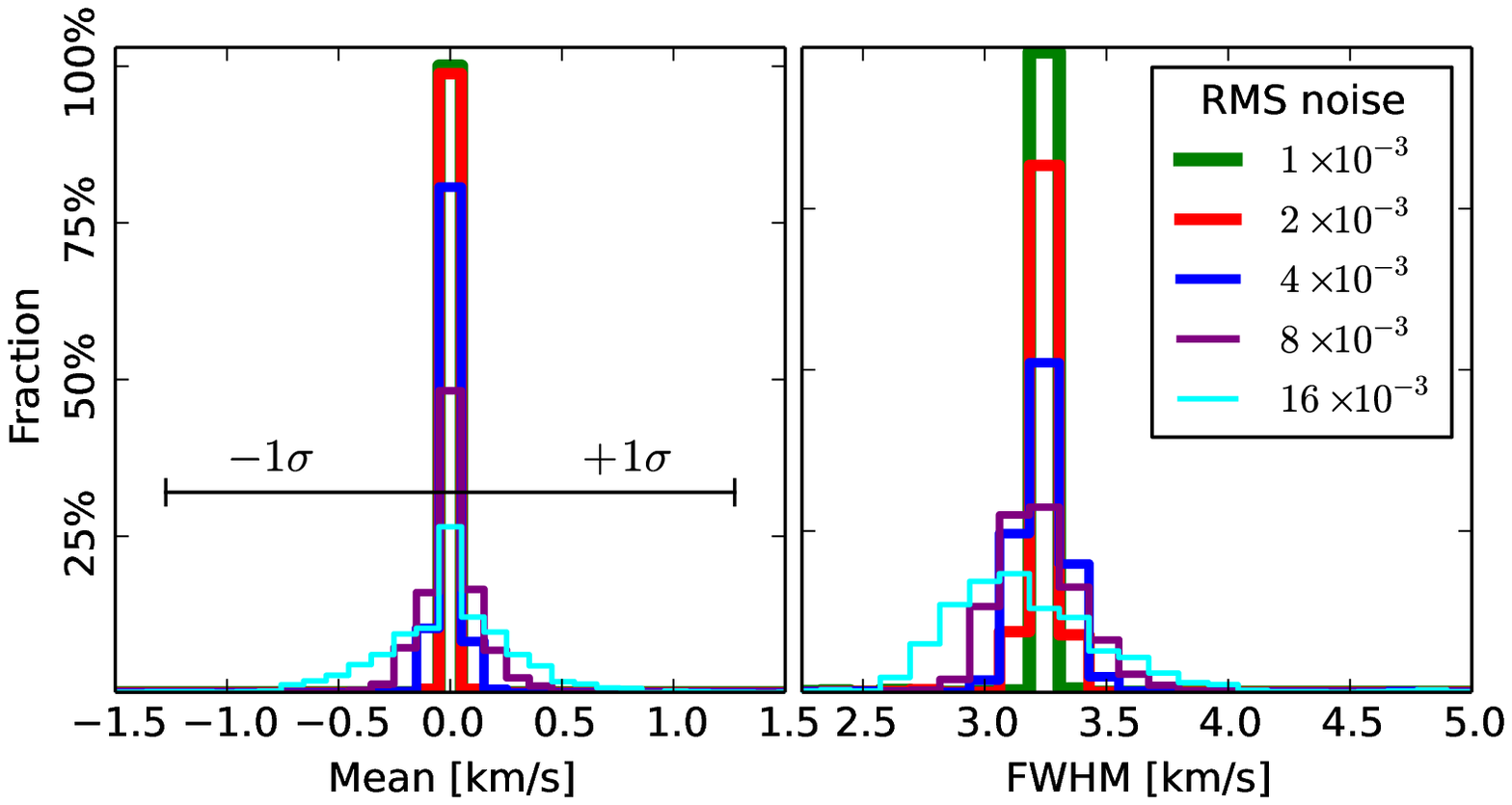}
    \caption{Monte Carlo test of AGD robustness to increasing noise.  
    The left and right panels show the distribution of guessed positions and 
    FWHMs, respectively, for injected components which all have a fixed
    shape of $a=0.1$, $\mu=0\,\rm km\,s^{-1}$, and ${\rm FWHM}=3\,\rm km\,s^{-1}$. Different line thicknesses 
    and colors represent different RMS noise values, ranging from 
    $1\times 10^{-3}$ to $16\times 10^{-3}$.  The horizontal bracket displays
    the $\pm 1\sigma $ width of the injected component.}
    \label{f-stability}
\end{figure}

Regularized derivatives (Section \ref{s-reg}) 
are insensitive to noise on spatial scales less than
that set by the regularization 
parameter $\log \alpha$ (Equation \ref{e-scale}).
Because the observational sensitivity of 21-SPONGE
data is uniform and very high, we next demonstrate
that AGD is robust to varying noise 
properties by characterizing the guessed position 
and FWHM of a Gaussian component 
with fixed shape in data with increasing noise intensity.  
Figure \ref{f-stability} shows that $\sim 100\%$ of 
component guesses remain within $\pm 1\sigma$ distance 
of the true component positions for noise intensities ranging from
1--$16\times {\rm RMS}$.  Over the same
range in noise, the guesses FWHMs varied by $\pm 20\%$.
Therefore, varying the noise properties has little
effect on AGDs performance, making AGD a robust tool to
analyze heterogeneous datasets with varying sensitivities.

\section{Discussion and Conclusions}
\label{s-conclude}
We have presented an algorithm, named
Autonomous Gaussian Decomposition (AGD), which 
produces optimized initial guesses for the parameters
of a multi-component Gaussian fit to spectral data.  
AGD uses derivative spectroscopy and bases its 
guesses on the properties of the first four numerical 
derivatives of the data.  The numerical derivatives 
are calculated using regularized optimization, 
and the smoothness of the derivatives is controlled by
the regularization parameter $\log \alpha$.  
Superivsed machine learning is then used 
to train the algorithm to choose the optimal value of
$\log \alpha$ which maximizes the accuracy of component
identification on a given training set of data
with known decompositions.

We test AGD by comparing its results to human-derived
Gaussian decompositions for 21 spectra from the 21-SPONGE
survey (Murray et al. 2014, in prep.).  For this test, we train
the algorithm on results from the independent 
Millenium survey \citep{heiles2003}.
We find that AGD performs comparably to humans when 
decomposing spectra in terms of number of components guessed, 
the residuals in the resulting fit, and the shape parameters
of the resulting components.  AGD's performance is
affected little by varying observational noise intensity 
until the point where components fall below the S/N 
threshold (i.e., completeness). Combined with Monte Carlo 
simulation, we use AGD to measure the H{\sc i} line 
completeness of 21-SPONGE data as a function of H{\sc i} 
peak optical depth and velocity width.
Thus, AGD is well suited for
helping to interpret the big spectral data incoming from
the SKA and SKA-pathfinder radio telescopes.

The time required for GaussPy to decompose
a spectrum  varies with
the number of channels and complexity of components.
For data consisting of between 100 to a few thousand channels,
and containing between 1 and 15 components, the
time required (for initial guesses + final fit) is
between 0.1 to a few seconds for each spectrum
on a single $3\,\rm GHz$ computer core.

AGD is distinct from Bayesian spectral
line finding algorithms \citep[e.g., ][]{allison2011} in 
terms of the criteria used in deciding the number of components.
Where the Bayesian approach chooses the number based on
the Bayesian evidence, AGD uses machine learning 
and is motivated by the answers
in the training set.  This machine learning
approach requires one to produce a training
set, but allows for more flexibility in telling the
algorithm how 
spectra should be decomposed.

In Section \ref{s:performance}, we used AGD to decompose 
spectra into Gaussian components
which correspond to physical clouds in the ISM. 
However, AGD can provide a useful parametrization of 
spectral data even when there is no {\em physical} 
motivation to represent the data as independent Gaussian functions.
For example, AGD could potentially be used to compress
the data volume of wide-bandwidth spectra for easy 
data transportation, or on-the-fly viewing.  
For example, If a $16\times 10^{3}$ channel spectrum contains signals 
which can be represented by $\sim 10$ Gaussian components, 
then by recasting the data\footnote{The ASKAP spectrometer provides
a total of 16384 channels} into  Gaussian component 
lists one could achieve a data compression 
factor of $\sim 500$.

\acknowledgements
This work was supported by the NSF Early Career
Development (CAREER) Award AST-1056780 and
by NSF Grant No. 1211258.
C. M. acknowledges support by the National Science 
Foundation Graduate Research Fellowship and the Wisconsin
Space Grant Institution.  We thank Elijah Bernstein-Cooper,
Matthew Turk, and James Foster for useful discussions.
We thank Lauren Michael and the University of 
Wisconsin's Center for High-Throughput computing 
for their help and support with HTCondor. 
CV-C expresses his appreciation towards
the Aspen Center for Physics for their hospitality.
The National Radio Astronomy Observatory 
is a facility of the National Science Foundation operated 
under cooperative agreement by Associated 
Universities, Inc.
The Arecibo Observatory is operated by
SRI International under a cooperative agreement with
the National Science Foundation (AST-1100968), and
in alliance with Ana G. M\'endez-Universidad
Metropolitana, and the Universities 
Space Research Association.

\appendix

\section{Deblended Amplitude Guesses}
\label{a-deblend}
AGD ``de-blends'' the naive amplitude guesses using the 
fact that when the parameters $\sigma_n$ and $\mu_n$ are
fixed, the multi-component Gaussian model becomes a {\em linear} 
function of the component amplitudes.
Therefore, the naive amplitude estimates can be written as
a linear combination of true deblended 
amplitudes $a^{\rm true}$, weighted by the overlap 
from each neighboring component. 
This system of linear equations is expressed in matrix 
form \citep[see, e.g., ][]{Kurczynski2010} as
\begin{equation}
     \left(
     \begin{matrix}
     B_{11} & \cdots & B_{1N} \\
     \vdots & \ddots & \vdots \\
     B_{N1} & \cdots & B_{NN}
     \end{matrix} \right)     
     \left( \begin{matrix}
     a^{\rm true}_{1} \\
     \vdots \\
          a^{\rm true}_{N}
     \end{matrix} \right)
      =
      \left(\begin{matrix}
      a^{\rm naive}_{1} \\
     \vdots \\
          a^{\rm naive}_{N}
      \end{matrix}\right)
\end{equation}
where
\begin{equation}
    B_{ij} = e^{  \frac{ -(\mu_i-\mu_j)^2} {2  \sigma_j^2}}.
\end{equation}
The elements of matrix $B_{ij}$ represent the overlap of
component $j$ onto the center of component $i$.  
When components are negligibly blended, 
$B_{ij}$ is equal to the identity matrix and 
$a_n^{\rm true}=a_n^{\rm naive}$.
The ``true'' de-blended amplitude estimates $a^{\rm true}_n$
are then found using the normal equations of linear least squares minimization
to be
\begin{equation}
    a^{\rm true} = (B^{\rm T}B)^{-1}B^{\rm T}a^{\rm naive}.
\end{equation} 
In practice, we compute the
solution for $a^{\rm true}$ through numerical optimization
to avoid inverting a possibly singular matrix $B$.
If all the de-blended amplitude estimates are greater 
than zero (physically valid), then they are adopted as 
the amplitude guesses; if any are $\leq 0$ (caused by 
errors in the estimates of $\mu_n$, $\sigma_n$, or the number
of components), the naive amplitudes are retained.  Therefore, 
\begin{equation}
a_n = 
\begin{cases}
a_n^{\rm true} & \mbox{if all }a_n^{\rm true}>0 \\
a_n^{\rm naive} & \rm otherwise.
\end{cases}
\end{equation}

\section{Momentum-driven Gradient Descent}
\label{a-gd}
The regularization parameter $\log \alpha$ (which is
generally a multi-dimensional vector; see, e.g., 
Appendix \ref{a-two-phase}) is tuned to
maximize the accuracy of component guesses 
(Equation \ref{e-acc}) using gradient descent
with momentum.  We define the cost
function $J$ that we wish to minimize in order to
find this solution as 
\begin{equation}
J(\log \alpha) = - {\rm ln}\, {\mathcal A(\log \alpha)}.
\end{equation}

In traditional gradient descent, updates to the 
parameter vector $\log \alpha$ are made by moving in 
the direction of greatest decrease in the cost function, 
i.e., $\Delta \log \alpha = - \lambda \nabla J(\log \alpha)$, and
the learning rate $\lambda$ controls the step size.
Our cost function $J(\log\alpha)$ is 
highly non-convex, so we use gradient 
descent \citep[see, e.g.,][]{press1992} 
with added momentum to push through local 
noise valleys.  Therefore, at the $n^{\rm th}$ iteration, our 
parameter update is given by 
\begin{equation}
\label{e-update}
  \Delta \log \alpha_n = -\lambda \nabla J(\log \alpha_n) + \phi \Delta \log \alpha_{n-1},
\end{equation}
where the ``momentum'' $\phi$ controls the degree to which the
previous update influences the current update.  

Because the decision function (i.e., 
Equations \ref{e-d1}--\ref{e-d3}) representing the 
success or failure for individual component guesses is 
binary in nature, the cost function $J(\log \alpha)$ is a 
piecewise-constant surface on small scales (see Figure \ref{f-tracks}).  
Therefore, 
in order to probe the large-scale slope of the cost function
surface, we use a relatively large value for the finite
difference step size when computing the gradient.  
For example,  the $i_{th}$ component of
the gradient in  Equation \ref{e-update} is defined according to
\begin{equation}
\nabla_i J(\log\alpha) = \frac{J(\log \alpha_i+\epsilon)-J(\log \alpha_i-\epsilon)}{2\epsilon},
\end{equation}
where $\epsilon$ is the finite-difference step size which we set
to $\epsilon=0.25$.  Figure \ref{f-tracks} shows example tracks
of $\log \alpha=(\log \alpha_1, \log \alpha_2)$ when using gradient descent with momentum
during AGD's two-phase training on the 21-SPONGE data. We find that 
small-scale local optima are ignored effectively during the search
for large-scale optima.

\section{Two-phase Gaussian Decomposition}
\label{a-two-phase}

Two-phase decompositions allow researchers to decompose
spectra which contain components that are drawn from 
two distributions with very different widths.  
GaussPy performs two-phase decomposition by first 
applying the usual AGD algorithm but with  
a non-zero threshold used in Equation \ref{e-c2}: ${\rm d}f^2/{{\rm d} x^2} < e_2$,
which locates only the narrowest components in the
data so that they can be removed.
The parameters of just these narrow components are next found
by minimizing the sum of squared residuals $K$ between the 
second derivative of the data and the second derivative 
of a model consisting of only these narrow 
components, $\{a_n^{\mathcal N},\mu_n^{\mathcal N}, \sigma_n^{\mathcal N}\}_{n=1}^{N}\equiv \mathcal N$,
given by
\begin{equation}
K({\mathcal N}) = \sum_{x}\left|  
\frac{\rm d^2}{{\rm d}x^2}f(x) - \frac{\rm d^2}{{\rm d}x^2}\sum_{n} G(x;a_n^{\mathcal N},\mu_n^{\mathcal N}, \sigma_n^{\mathcal N})\right|^2.
\end{equation}
The narrow components are fit to the data on the basis of 
their second derivatives so
that the signals from wider components, which they may be 
superposed on, are attenuated by a factor
$\sim  \sigma_{\rm narrow}^2 / \sigma_{\rm broad} ^2$.
The residual spectrum is
then fed back into AGD to search for broader components using
a larger value of $\log \alpha$ and 
setting $e_2=0$.

\bibliographystyle{apj}

\begin{thebibliography}{}
\expandafter\ifx\csname natexlab\endcsname\relax\def\natexlab#1{#1}\fi

\bibitem[{{Allison} {et~al.}(2012){Allison}, {Sadler}, \&
  {Whiting}}]{allison2011}
{Allison}, J.~R., {Sadler}, E.~M., \& {Whiting}, M.~T. 2012, \pasa, 29, 221

\bibitem[{{Audit} \& {Hennebelle}(2005)}]{audit2005}
{Audit}, E., \& {Hennebelle}, P. 2005, \aap, 433, 1

\bibitem[{{Begum} {et~al.}(2010){Begum}, {Stanimirovi{\'c}}, {Goss}, {Heiles},
  {Pavkovich}, \& {Hennebelle}}]{begum2010}
{Begum}, A., {Stanimirovi{\'c}}, S., {Goss}, W.~M., {et~al.} 2010, \apj, 725,
  1779

\bibitem[{Bishop(2006)}]{bishop2006}
Bishop, C.~M. 2006, Pattern Recognition and Machine Learning (Information
  Science and Statistics) (Secaucus, NJ, USA: Springer-Verlag New York, Inc.)

\bibitem[{{Chartrand}(2011)}]{chartrand2011}
{Chartrand}, R. 2011, ISRN Applied Mathematics, 2011, doi:10.5402/2011/164564

\bibitem[{{Clark} {et~al.}(2012){Clark}, {Glover}, {Klessen}, \&
  {Bonnell}}]{clark2012}
{Clark}, P.~C., {Glover}, S.~C.~O., {Klessen}, R.~S., \& {Bonnell}, I.~A. 2012,
  \mnras, 424, 2599

\bibitem[{{Crovisier} {et~al.}(1980){Crovisier}, {Kazes}, \&
  {Aubry}}]{crovisier1980}
{Crovisier}, J., {Kazes}, I., \& {Aubry}, D. 1980, \aaps, 41, 229

\bibitem[{Dean \& Ghemawat(2004)}]{dean2004}
Dean, J., \& Ghemawat, S. 2004, in Proceedings of the 6th Conference on
  Symposium on Opearting Systems Design \& Implementation - Volume 6, OSDI'04
  (Berkeley, CA, USA: USENIX Association), 10--10

\bibitem[{{Dickey} {et~al.}(2003){Dickey}, {McClure-Griffiths}, {Gaensler}, \&
  {Green}}]{dickey2003}
{Dickey}, J.~M., {McClure-Griffiths}, N.~M., {Gaensler}, B.~M., \& {Green},
  A.~J. 2003, \apj, 585, 801

\bibitem[{{Dickey} {et~al.}(1978){Dickey}, {Terzian}, \&
  {Salpeter}}]{dickey1978}
{Dickey}, J.~M., {Terzian}, Y., \& {Salpeter}, E.~E. 1978, \apjs, 36, 77

\bibitem[{Draine(2010)}]{draine2010}
Draine, B. 2010, Physics of the Interstellar and Intergalactic Medium,
  Princeton Series in Astrophysics (Princeton University Press)

\bibitem[{Fell(1983)}]{fell1983}
Fell, A.~F. 1983, TrAC Trends in Analytical Chemistry, 2, 63

\bibitem[{{Field} {et~al.}(1969){Field}, {Goldsmith}, \& {Habing}}]{field1969}
{Field}, G.~B., {Goldsmith}, D.~W., \& {Habing}, H.~J. 1969, \apjl, 155, L149

\bibitem[{{Haud}(2000)}]{haud2000}
{Haud}, U. 2000, \aap, 364, 83

\bibitem[{{Heiles}(2001)}]{heiles2001}
{Heiles}, C. 2001, \apjl, 551, L105

\bibitem[{{Heiles} \& {Troland}(2003)}]{heiles2003}
{Heiles}, C., \& {Troland}, T.~H. 2003, \apjs, 145, 329

\bibitem[{{Heitsch} {et~al.}(2005){Heitsch}, {Burkert}, {Hartmann}, {Slyz}, \&
  {Devriendt}}]{heitsch2005}
{Heitsch}, F., {Burkert}, A., {Hartmann}, L.~W., {Slyz}, A.~D., \& {Devriendt},
  J.~E.~G. 2005, \apjl, 633, L113

\bibitem[{{Hennebelle} \& {Iffrig}(2014)}]{hennebelle2014}
{Hennebelle}, P., \& {Iffrig}, O. 2014, ArXiv e-prints, arXiv:1405.7819

\bibitem[{Hunter(2007)}]{hunter2007}
Hunter, J.~D. 2007, Computing In Science \& Engineering, 9, 90

\bibitem[{Ivezi\`c {et~al.}(2014)Ivezi\`c, Connolly, VanderPlas, \&
  Gray}]{zeljko2014}
Ivezi\`c, v., Connolly, A.~J., VanderPlas, J.~T., \& Gray, A. 2014, Statistics,
  Data Mining, and Machine Learning in Astronomy, stu - student edition edn.,
  Princeton Series in Modern Observational Astronomy (Princeton University
  Press), 544

\bibitem[{Jones {et~al.}(2001--)Jones, Oliphant, Peterson,
  {et~al.}}]{jones2001}
Jones, E., Oliphant, T., Peterson, P., {et~al.} 2001--, {SciPy}: Open source
  scientific tools for {Python}, [Online; accessed 2014-08-20]

\bibitem[{{Kalberla} {et~al.}(2005){Kalberla}, {Burton}, {Hartmann}, {Arnal},
  {Bajaja}, {Morras}, \& {P{\"o}ppel}}]{kalberla2005}
{Kalberla}, P.~M.~W., {Burton}, W.~B., {Hartmann}, D., {et~al.} 2005, \aap,
  440, 775

\bibitem[{{Kalberla} {et~al.}(1985){Kalberla}, {Schwarz}, \&
  {Goss}}]{kalberla1985}
{Kalberla}, P.~M.~W., {Schwarz}, U.~J., \& {Goss}, W.~M. 1985, \aap, 144, 27

\bibitem[{{Kr{\v c}o} {et~al.}(2008){Kr{\v c}o}, {Goldsmith}, {Brown}, \&
  {Li}}]{krco2008}
{Kr{\v c}o}, M., {Goldsmith}, P.~F., {Brown}, R.~L., \& {Li}, D. 2008, \apj,
  689, 276

\bibitem[{{Kurczynski} \& {Gawiser}(2010)}]{Kurczynski2010}
{Kurczynski}, P., \& {Gawiser}, E. 2010, \aj, 139, 1592

\bibitem[{Levenberg(1944)}]{levenberg1944}
Levenberg, K. 1944, Quarterly of Applied Mathematics, 2, 164

\bibitem[{{Liszt}(2001)}]{liszt2001}
{Liszt}, H. 2001, \aap, 371, 698

\bibitem[{{Mac Low} {et~al.}(2005){Mac Low}, {Balsara}, {Kim}, \& {de
  Avillez}}]{maclow2005}
{Mac Low}, M.-M., {Balsara}, D.~S., {Kim}, J., \& {de Avillez}, M.~A. 2005,
  \apj, 626, 864

\bibitem[{{McKee} \& {Ostriker}(1977)}]{mckee1977}
{McKee}, C.~F., \& {Ostriker}, J.~P. 1977, \apj, 218, 148

\bibitem[{{Mebold} {et~al.}(1982){Mebold}, {Winnberg}, {Kalberla}, \&
  {Goss}}]{mebold1982}
{Mebold}, U., {Winnberg}, A., {Kalberla}, P.~M.~W., \& {Goss}, W.~M. 1982,
  \aap, 115, 223

\bibitem[{{Murray} {et~al.}(2014){Murray}, {Lindner}, {Stanimirovi{\'c}},
  {Goss}, {Heiles}, {Dickey}, {Pingel}, {Lawrence}, {Jencson}, {Babler}, \&
  {Hennebelle}}]{murray2014}
{Murray}, C.~E., {Lindner}, R.~R., {Stanimirovi{\'c}}, S., {et~al.} 2014,
  \apjl, 781, L41

\bibitem[{{Nidever} {et~al.}(2008){Nidever}, {Majewski}, \&
  {Burton}}]{nidever2008}
{Nidever}, D.~L., {Majewski}, S.~R., \& {Burton}, W.~B. 2008, \apj, 679, 432

\bibitem[{Nocedal \& Wright(2006)}]{nocedal2006}
Nocedal, J., \& Wright, S. 2006, Numerical Optimization, Springer Series in
  Operations Research and Financial Engineering (Springer)

\bibitem[{{Press} {et~al.}(1992){Press}, {Teukolsky}, {Vetterling}, \&
  {Flannery}}]{press1992}
{Press}, W.~H., {Teukolsky}, S.~A., {Vetterling}, W.~T., \& {Flannery}, B.~P.
  1992, {Numerical recipes in FORTRAN. The art of scientific computing}

\bibitem[{{Roy} {et~al.}(2013){Roy}, {Kanekar}, \& {Chengalur}}]{roy2013}
{Roy}, N., {Kanekar}, N., \& {Chengalur}, J.~N. 2013, \mnras, 436, 2366

\bibitem[{Skilling(2004)}]{skilling2004}
Skilling, J. 2004, AIP Conference Proceedings, 735, 395

\bibitem[{{Stutzki} \& {Guesten}(1990)}]{stutzki1990}
{Stutzki}, J., \& {Guesten}, R. 1990, \apj, 356, 513

\bibitem[{{Taylor} {et~al.}(1999){Taylor}, {Carilli}, \& {Perley}}]{taylor1999}
{Taylor}, G.~B., {Carilli}, C.~L., \& {Perley}, R.~A., eds. 1999, Astronomical
  Society of the Pacific Conference Series, Vol. 180, {Synthesis Imaging in
  Radio Astronomy II}

\bibitem[{{Tikhonov}(1963)}]{tikhonov1963}
{Tikhonov}, A.~N. 1963, Soviet Mathematics-Doklady, 4, 1624

\bibitem[{{Vogel}(2002)}]{vogel2002}
{Vogel}, C.~R. 2002, {Computational Methods for Inverse Problems}, Vol.~23
  (Frontiers in Applied Mathematics: Society for Industrial and Applied
  Mathematics SIAM)

\bibitem[{{Wallington} {et~al.}(1994){Wallington}, {Narayan}, \&
  {Kochanek}}]{wallington1994}
{Wallington}, S., {Narayan}, R., \& {Kochanek}, C.~S. 1994, \apj, 426, 60

\bibitem[{Walt {et~al.}(2011)Walt, Colbert, \& Varoquaux}]{walt2011}
Walt, S. v.~d., Colbert, S.~C., \& Varoquaux, G. 2011, Computing in Science \&
  Engineering, 13, 22

\bibitem[{{Westerlund} \& {Harris}(2014)}]{westerlund2014}
{Westerlund}, S., \& {Harris}, C. 2014, ArXiv e-prints, arXiv:1407.4958

\bibitem[{{Whiting}(2012)}]{whiting2012}
{Whiting}, M.~T. 2012, \mnras, 421, 3242

\bibitem[{{Williams} {et~al.}(1994){Williams}, {de Geus}, \&
  {Blitz}}]{williams1994}
{Williams}, J.~P., {de Geus}, E.~J., \& {Blitz}, L. 1994, \apj, 428, 693

\bibitem[{{Wolfire} {et~al.}(2003){Wolfire}, {McKee}, {Hollenbach}, \&
  {Tielens}}]{wolfire2003}
{Wolfire}, M.~G., {McKee}, C.~F., {Hollenbach}, D., \& {Tielens}, A.~G.~G.~M.
  2003, \apj, 587, 278

\bibitem[{Zhu {et~al.}(1997)Zhu, Byrd, Lu, \& Nocedal}]{zhu1997}
Zhu, C., Byrd, R.~H., Lu, P., \& Nocedal, J. 1997, ACM Trans. Math. Softw., 23,
  550

\end{thebibliography}

\end{document}